\documentclass{revtex4}
\usepackage{graphicx}
\usepackage{amsmath}
\usepackage{amsfonts}
\usepackage{upgreek}
\usepackage{slashed}
\usepackage{color}
\usepackage{mathrsfs}
\DeclareMathOperator{\thetaf}{\uptheta}
\DeclareMathOperator{\Tr}{Tr}

\begin{document}
\title{Fermion Self-energy and Pseudovector Condensate in NJL Model with External Magnetic Field}
\author{Song Shi$^{1,2}$}
\author{Yi-Lun Du$^2$}
\author{Yi Tang$^3$}
\author{Yong-Hui Xia$^2$}
\author{Hong-Shi Zong$^{2,4}$}\email[]{zonghs@nju.edu.cn}
\affiliation{$^1$ Department of Physics, National University of Defense Technology, Changsha 410000, China}
\affiliation{$^2$ Department of Physics, Nanjing University, Nanjing 210093, China}
\affiliation{$^3$ Officers College of PAP, Chendu 610000, China}
\affiliation{$^4$ Joint Center for Particle, Nuclear Physics and Cosmology, Nanjing 210093, China}

\begin{abstract}
In this paper, we aim to study the complete self-energy in the fermion propagator within two-flavor NJL model in the case of finite temperature, chemical potential and external magnetic field. Through Fierz transformation we prove that the self-energy is not simply proportional to dynamical mass in the presence of chemical potential, moreover, it contains four kinds of condensates after introducing external magnetic field. We find out the appropriate and complete form of self-energy and establish new gap equations. We take two of the four condensates (scalar and pseudovector condensates) to make an approximation and simplify the gap equations. The numerical results show that the dynamical mass get a small quantitative modification after introducing pseudovector condensate comparing to classic result, the main properties of Nambu phase and Wigner phase are quite the same with classic ones. The pseudovector condensate also has a small gap between Nambu phase and Wigner phase, this provides us a new order parameter for the phase transition. We also find that pseudovector condensate could cause energy level splitting, this also provides a possibly observable property in astrophysics such as in magnetar.

\bigskip


\end{abstract}

\maketitle

\section{Introduction}
The phase structure of QCD matter has always been an important and attractive topic in theoretical physics \cite{fukushima,costa,chao,zhao,jiang,xu,fu}. In relativistic heavy-ion collisions, the produced QCD matter will go though a phase transition or a crossover as time goes by.  Either way, the state of QCD matter is believed to change from quark-gluon plasma to hadronic matter in this process. Its physical properties and dynamical behaviors such as chiral symmetry and confinement are altered along with the change of the state.

At the early stage of noncentral collision, the QCD matter produces extremely strong magnetic field \cite{kharzeev}, which brings about obvious magnetic effects. Therefore studying QCD matter's properties under the influence of magnetic field becomes a meaningful and important subject. So far, many relevant theories and models have been proposed and it is shown that the quark condensate are strengthened by magnetic field, which is known as `Magnetic Catalysis' \cite{gusynin,gusynin2,ebert,shov,alan}. Consequently, the QCD phase diagram is related to magnetic field \cite{andersen,endrodi}.

NJL model is quite a useful and convenient tool to qualitatively study QCD matter states \cite{ghosh,menezes,menezes2,shi,du,cui,du2,lu,fukushima,costa}. For a NJL model we usually apply mean field approximation to deal with the four fermion interaction terms, namely, $(\bar\psi\psi)^2\to2\langle\bar\psi\psi\rangle(\bar\psi\psi)-\langle\bar\psi\psi\rangle^2$, $(i\bar\psi\gamma^5\vec\tau\psi)^2\to2\langle i\bar\psi\gamma^5\vec\tau\psi\rangle\cdot(i\bar\psi\gamma^5\vec\tau\psi)-\langle i\bar\psi\gamma^5\vec\tau\psi\rangle^2$. It is believed that this approximation is equivalent to Dyson-Schwinger equations with contact interaction treatment, hence the gap equation can be written as
\begin{equation}
 \frac{\Sigma}{G}\int d^4x=i\int d^4x\,\langle x|\gamma^\mu\hat{S}\gamma_\mu|x\rangle,\label{gap}
\end{equation}
\begin{equation}
 \hat{S}^{-1}=\slashed{p}-m-\Sigma,
\end{equation}
\begin{equation}
 \Sigma=\sigma+i\gamma^5\vec\pi\cdot\vec\tau,\qquad\sigma=-\frac{G}{N_\text{c}}\langle\bar\psi\psi\rangle,\qquad\vec\pi=-\frac{G}{N_\text{c}}\langle i\bar\psi\gamma^5\vec\tau\psi\rangle.
\end{equation}

In the above equations, $\Sigma$ represents self-energy of fermion propagator, it contains dynamical mass $\sigma$, which is generated by non-perturbative effect, more specifically, dynamical chiral symmetry breaking. Generally speaking, there is $\vec\pi=0$ in Eq. (\ref{gap}), which leads to $\Sigma=\sigma$. Therefore it is usually more convenient to study dynamical mass directly rather than discuss a general form of self-energy. But in M. Asakawa and K. Yazaki's work \cite{masayuki}, they have pointed that, in a self-consistent mean-field approximation, the self-energy does not simply equal dynamical mass, which reveals with the help of the Fierz transformation. When chemical potential $\mu$ is not zero, the actual self-energy should be written as $\Sigma=\sigma+a\gamma^0$ to guarantee the self-consistency of gap equation (\ref{gap}) . In this new $\Sigma$, we can combine $a$ with chemical potential as a renormalized chemical potential $\mu_\text{r}=\mu-a$.

In this paper, we are about to study the self-energy problem in two flavor NJL model with temperature, chemical potential and external magnetic field, the self-energy must not simply equal dynamical mass. In order to find out the appropriate self-energy, we start from the most general form, a $(4\times4)_\text{s}\otimes(2\times2)_\text{f}$ matrix (spinor space and flavor space), and rule out its inappropriate parts. The detail is discussed in the beginning of section II. In section II, we give a simple deduction of gap equation (the detailed deduction is shown in Appendix \ref{sec:append}) and analyse numerical results. Section III is our conclusion.

\section{The Gap Equations and Numerical Results}
\subsection{Selfenergy of NJL Model and Gap Equations}
A two-flavor NJL model Lagrangian with external magnetic field in Minkowski space is
\begin{equation}
 \mathcal{L}=\bar\psi(i\slashed{\partial}+e\slashed{A}\otimes Q)\psi+G[(\bar\psi\psi)^2+(i\bar\psi\gamma^5\otimes\vec\tau\psi)^2],\label{njl}
\end{equation}
\begin{equation}
 (A_0, A_1, A_2, A_3)=(0, \frac{B}{2}x^2, -\frac{B}{2}x^1, 0),\label{mag}
\end{equation}
\begin{equation}
 Q=\left(\begin{array}{cc}q_\text{u}&0\\0&q_\text{d}\end{array}\right),\qquad q_\text{u}=\frac{2}{3},\qquad q_\text{d}=-\frac{1}{3},
\end{equation}

According to Ref. \cite{masayuki}, if one wants to apply mean field approximation to Eq. (\ref{njl}), applying mean field approximation to $(\bar\psi\psi)$ and $(i\bar\psi\gamma^5\otimes\vec\tau\psi)$ is not enough. Through Fierz transformation, the interaction terms in Lagrangian produce more four fermions interaction terms, and now we can apply mean field approximation.

Let $\mathcal{L}_\text{I}$ represent the interaction terms of four fermions in Eq. (\ref{njl}),
\begin{equation}
 \mathcal{L}_\text{I}=G[(\bar\psi\psi)^2+(i\bar\psi\gamma^5\vec\tau\psi)^2],\label{int1}
\end{equation}
and Fierz transformation of $\mathcal{L}_\text{I}$ yields \cite{klev}
\begin{equation}
 \begin{split}
  \mathscr{F}(\mathcal{L}_\text{I})=&\frac{G}{4N_\text{c}}[(\bar\psi\psi)^2+(i\bar\psi\gamma^5\vec\tau\psi)^2-(\bar\psi\vec\tau\psi)^2-(i\bar\psi\gamma^5\psi)^2\\
  &-2(\bar\psi\gamma^\mu\psi)^2-2(\bar\psi\gamma^5\gamma^\mu\psi)^2+(\bar\psi\sigma^{\mu\nu}\psi)^2-(\bar\psi\sigma^{\mu\nu}\vec\tau\psi)^2].
 \end{split}\label{int2}
\end{equation}

In Klevansky's article \cite{klev}, he had mentioned that there are three equivalent four fermions interaction terms by using Fierz transformation(see Eq. (2.57) in his article, ),
\begin{equation}
 \mathcal{L}_\text{I}=G[(\bar\psi\psi)^2+(i\bar\psi\gamma^5\vec\tau\psi)^2],\label{int01}
\end{equation}
\begin{equation}
 \mathscr{F}(\mathcal{L}_\text{I}),\label{int02}
\end{equation}
\begin{equation}
 \frac{1}{2}[\mathcal{L}_\text{I}+\mathscr{F}(\mathcal{L}_\text{I})].\label{int03}
\end{equation}
The dynamic properties of these three interaction terms should be equivalent in a external-field-free NJL model, but in this paper, the presence of external magnetic field breaks their equivalence. Thus we have to decide which one of the three should be appropriate interaction terms, it seems Eq. (\ref{int03}) is an appropriate one for three reasons, first of all, Eq. (\ref{int03}) provides us more structures than Eq. (\ref{int01}), secondly, Eq. (\ref{int03}) is the only one that is obviously Fierz transformation invariant comparing to the other two, thirdly, Eq. (\ref{int03}) has $O(G)$ terms and $O(\frac{O}{4N_\text{c}})$ terms, while Eq. (\ref{int03}) only has $O(\frac{O}{4N_\text{c}})$ terms, it seems $O(G)$ might have dominant effect in dynamic process, hence Eq. (\ref{int03}) is quite more convincible than Eq. (\ref{int02}). In a more explicit form, Eq. (\ref{int03}) is
\begin{equation}
 \begin{split}
  \mathcal{L}_\text{I}'=&\frac{G}{2}(1+\frac{1}{4N_\text{c}})[(\bar\psi\psi)^2+(i\bar\psi\gamma^5\vec\tau\psi)^2]-\frac{G}{8N_\text{c}}[(\bar\psi\vec\tau\psi)^2+(i\bar\psi\gamma^5\psi)^2\\
  &+2(\bar\psi\gamma^\mu\psi)^2+2(\bar\psi\gamma^5\gamma^\mu\psi)^2-(\bar\psi\sigma^{\mu\nu}\psi)^2+(\bar\psi\sigma^{\mu\nu}\vec\tau\psi)^2].\label{int04}
 \end{split}
\end{equation}

Now applying mean field approximation to Eq. (\ref{int04}), we are led to a complex self-energy $\Sigma_\text{sf}$, which should be a $(4\times4)_\text{s}\otimes(2\times2)_\text{f}$ matrix
\begin{equation}
 \mathcal{L}_\text{I}'\to\mathcal{L}_\text{mean}=-\bar\psi\Sigma_\text{sf}\psi+\mathcal{L}_\text{M},\label{if}
\end{equation}
\begin{equation}
 \begin{split}
  \Sigma_\text{sf}=&-G(1+\frac{1}{4N_\text{c}})[\langle\bar\psi\psi\rangle+\langle i\bar\psi\gamma^5\vec\tau\psi\rangle\cdot(i\gamma^5\vec\tau)]+\frac{G}{4N_\text{c}}[\langle\bar\psi\vec\tau\psi\rangle\cdot\vec\tau+\langle i\bar\psi\gamma^5\psi\rangle i\gamma^5\\
  &+2\langle\bar\psi\gamma_\mu\psi\rangle\gamma^\mu+2\langle\bar\psi\gamma^5\gamma_\mu\psi\rangle(\gamma^5\gamma^\mu)
  -\langle\bar\psi\sigma_{\mu\nu}\psi\rangle\sigma^{\mu\nu}+\langle\bar\psi\sigma_{\mu\nu}\vec\tau\psi\rangle\cdot(\sigma^{\mu\nu}\vec\tau)].
 \end{split}\label{int3}
\end{equation}
consequently, the new Lagrangian is
\begin{equation}
 \mathcal{L}'=\bar\psi(/\kern-0.55em\hat{\Pi}-\Sigma_\text{sf})\psi+\mathcal{L}_\text{M},\label{new}
\end{equation}
\begin{equation}
 \hat\Pi_\mu=i\partial_\mu+eA_\mu\otimes Q,
\end{equation}
of course, in order to separate flavor space, one can also define
\begin{equation}
 \hat\Pi_\mu^\text{f}=\hat p_\mu+q_\text{f}eA_\mu,\qquad\text{f}=\text{u},\text{d},\qquad q_\text{f}=q_\text{u},q_\text{d}.
\end{equation}
$\mathcal{L}_\text{M}$ in Eq. (\ref{if}) is the sum of mean field square terms such as $\langle\bar\psi\psi\rangle^2$, $\langle\bar\psi\gamma^\mu\psi\rangle^2$ , we leave the detailed expression of $\mathcal{L}_\text{M}$ after simplifying $\Sigma_\text{sf}$.

Through the discussion of Ref. \cite{shi}'s Appendix B, we can safely assume that self-energy can firstly be simplify to $\Sigma_\text{sf}=\Sigma\otimes I_2$, here $\Sigma$ is a linear combination of $16$ Dirac matrices $\{I_4, \gamma^\mu, \gamma^5, \gamma^5\gamma^\mu, \sigma^{\mu\nu}\}$. Secondly, in Eq. (\ref{new}),  despite mean field approximation $\mathcal{L}'$ should still preserve the same Lorentz invariance as $\mathcal{L}$ does, and the presence of magnetic field $A_\mu$ degenerate the usual Lorentz invariance in $3+1$ dimension to $O(2)$ invariance in $x^1$-$x^2$ plane, therefore $\gamma^5\gamma^{1,2}$, $\gamma^5\gamma^{1,2}$, $\sigma^{23}$ and $\sigma^{31}$ should couple with $\hat\Pi_i^\text{f}$ in $\Sigma$ due to the requirement of covariation in $x^1$-$x^2$ plane, but these kinds of couplings conflict with mean field approximation in which $\Sigma$ must be a constant matrix, thus we conclude that $\Sigma$ does not have $\gamma^{1,2}$, $\gamma^5\gamma^{1,2}$, $\sigma^{23}$ and $\sigma^{31}$ as its components. Thirdly we expect $\Sigma_\text{sf}$ obeys parity symmetry, while the terms with $\langle\bar\psi\gamma^3\psi\rangle$, $\langle i\bar\psi\gamma^5\psi\rangle$, $\langle\bar\psi\sigma^{03}\psi\rangle$ in Eq. (\ref{int3}) violate parity, they are not allowed neither. Fourthly, the term with $\sigma^{12}$, it is believed that $\sigma^{12}$ should also couple with $\hat\Pi_1^\text{f}\hat\Pi_2^\text{f}$ to preserve $O(2)$ symmetry in $x^1$-$x^2$ plane, but beware that $\hat\Pi_1^\text{f}$ and $\hat\Pi_2^\text{f}$ are not commutable, hence it is legitimate having the terms like $(\hat\Pi_1^\text{f}\hat\Pi_2^\text{f}\sigma^{12}+\hat\Pi_2^\text{f}\hat\Pi_1^\text{f}\sigma^{21})$ ($\hat\Pi_1^\text{f}\hat\Pi_2^\text{f}\sigma^{12}+\hat\Pi_2^\text{f}\hat\Pi_1^\text{f}\sigma^{21}
=[\hat\Pi_1^\text{f},\hat\Pi_2^\text{f}]\sigma^{12}=-iq_\text{f}eB\sigma^{12}$) in $\Sigma$, we can safely assume $\sigma^{12}$ couple with a constant (probably relate to $eB$). Based on the above discussion, the appropriate self-energy should be written as
\begin{equation}
 \Sigma=\sigma+a\gamma^0+b\gamma^5\gamma^3+c\sigma^{12}.\label{form}
\end{equation}

Comparing Eq. (\ref{form}) with Eq. (\ref{int3}), $\sigma$, $a$, $b$ and $c$ separately correspond to
\begin{equation}
 \sigma=-G(1+\frac{1}{4N_\text{c}})\langle\bar\psi\psi\rangle,\qquad a=\frac{G}{2N_\text{c}}\langle\bar\psi\gamma^0\psi\rangle,\qquad b=-\frac{G}{2N_\text{c}}\langle\bar\psi\gamma^5\gamma^3\psi\rangle,\qquad c=-\frac{G}{4N_\text{c}}\langle\bar\psi\sigma^{12}\psi\rangle,\label{mean}
\end{equation}
and now we are able to write down the explicit expression of $\mathcal{L}_\text{M}$,

\begin{equation}
 \mathcal{L}_\text{M}=-\frac{2N_\text{c}}{4N_\text{c}+1}\frac{1}{G}\sigma^2+\frac{N_\text{c}}{G}a^2-\frac{N_\text{c}}{G}b^2-\frac{2N_\text{c}}{G}c^2.
\end{equation}

As we can see, $\sigma$ is the dynamic mass, it represents quark condensate in quark matter. $a$ is a vector condensate, with finite chemical potential presenting, it can be seen as a modification to chemical potential. $b$ is the pseudovector condensate, this is the parameter we study in this article. $c$ is tensor condensate, it is generally a minor but nonzero quantity, in this article we treat it as zero to simplify calculation, in Appendix \ref{sec:append}, we explain the reason why we don't include this condensate in our study of dynamic mass generating.

In a thermal system described by NJL model, the existence of temperature $T$ and chemical potential $\mu$ does not change the structure of $\Sigma_\text{sf}$ in Eq. (\ref{form}). In Appendix \ref{sec:append}, we present the detail deduction of gap equations with finite temperature and chemical potential. Putting Eqs. (\ref{f1}) and (\ref{f3}) into Eqs. (\ref{sigmaf1}) and (\ref{bf3}) separately, one can transform the sum of all polynomials with $\omega_m$ into hyperbolic functions. Base on the equation (looking up detailed deduction in Refs. \cite{menezes,kapusta})
\begin{equation}
 \sum_m\ln\{\beta^2[(\omega_m+i\mu)^2+x^2]\}=\beta x+\ln[1+e^{-\beta(x-\mu)}]+\ln[1+e^{-\beta(x+\mu)}],\quad x\in\mathbb{R},
\end{equation}
we have the new gap equations described as below:
\begin{equation}
 \begin{split}
  \frac{2}{4N_\text{c}+1}\frac{4\pi^2}{G}=&\sum_\text{f}\frac{|q_\text{f}|eB}{\sqrt{\pi}}\int^{+\infty}_0\frac{\coth(|q_\text{f}|eBs)}{\sqrt{s}}\,ds
  \int\frac{1}{2}\bigg(\frac{\omega_{+}}{\omega}e^{-\omega^2_{+}s}+\frac{\omega_{-}}{\omega}e^{-\omega^2_{-}s}\bigg)\,dp_3\\
  &-eB\int\frac{1}{\omega}\bigg[\frac{1}{1+e^{\beta(\omega_{+}+\mu_\text{r})}}+\frac{1}{1+e^{\beta(\omega_{-}-\mu_\text{r})}}\bigg]\,dp_3\\
  &-\sum_\text{f}2|q_\text{f}|eB\int\frac{1}{\omega}\sum_{n=1}^{+\infty}(F_{+n\text{f}}+F_{-n\text{f}})\,dp_3
  +2eB\thetaf(|b|-\sigma)\ln\bigg(\frac{|b|+\sqrt{b^2-\sigma^2}}{\sigma}\bigg),
 \end{split}\label{sigma}
\end{equation}
\begin{equation}
 \begin{split}
  \frac{4\pi^2}{G}b=&\sum_\text{f}\frac{|q_\text{f}|eB}{2\sqrt{\pi}}\int_0^{+\infty}\frac{\coth(|q_\text{f}|eBs)}{\sqrt{s}}\,ds
  \int(\omega_{+}e^{-\omega^2_{+}s}-\omega_{-}e^{-\omega^2_{-}s})\,dp_3\\
  &+eB\int\bigg[\frac{1}{1+e^{\beta(\omega_{-}-\mu_\text{r})}}-\frac{1}{1+e^{\beta(\omega_{+}+\mu_\text{r})}}\bigg]\,dp_3\\
  &+\sum_\text{f}|q_\text{f}|eB\int\sum_{n=1}^{+\infty}(F_{-n\text{f}}-F_{+n\text{f}})\,dp_3+4eB[\thetaf(-b-\sigma)-\thetaf(b-\sigma)]\sqrt{b^2-\sigma^2},
 \end{split}\label{b}
\end{equation}
\begin{equation}
 F_{\pm n\text{f}}=\frac{\omega_\pm}{\omega_{\pm n\text{f}}}\frac{1}{1+e^{\beta(\omega_{\pm n\text{f}}-\mu_\text{r})}}
 +\frac{\omega_\pm}{\omega_{\pm n\text{f}}}\frac{1}{1+e^{\beta(\omega_{\pm n\text{f}}+\mu_\text{r})}},\quad n\in\mathbb{Z}^+. \label{Fn}
\end{equation}
the explicit expressions of $\omega_\pm$ and $\omega_{\pm n\text{f}}$ are shown in Appendix \ref{sec:append}.

Integrals of proper time `$s$' in Eqs. (\ref{sigma}) and (\ref{b}) need a cutoff. One can prove that when $T\to0$, $\mu_\text{r}=0$ and $b=0$, Eq. (\ref{sigma}) degenerates to the normal gap equation with external magnetic field
\begin{equation}
 \frac{2}{4N_\text{c}+1}\frac{4\pi^2}{G}=\sum_\text{f}|q_\text{f}|eB\int_0^{+\infty}\frac{e^{-\sigma^2s}}{s}\coth(|q_\text{f}|eBs)\,ds,
\end{equation}
this means we can apply the same regularization scheme \cite{shi,inagaki} to $s$ as normal gap equation does,
\begin{equation}
 \int_0^{+\infty}ds\to\int_{1/\Lambda^2}^{+\infty}ds,
\end{equation}
here the cutoff energy scale $\Lambda$ and coupling constant $G$ are assigned as
\begin{equation}
 \Lambda=0.991\text{GeV},\qquad G=\big(25.4\times\frac{2}{4N_\text{c}+1}\big)\text{GeV}^{-2}=3.91\text{GeV}^{-2}.
\end{equation}

\subsection{Numerical Results and Discussions}
\begin{figure}
 \centering
 \includegraphics[width=3in]{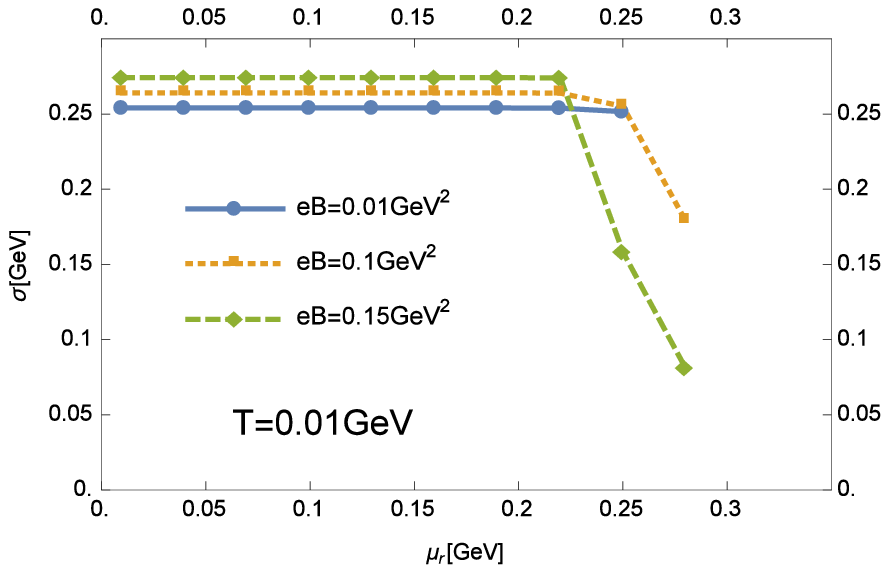}
 \includegraphics[width=3in]{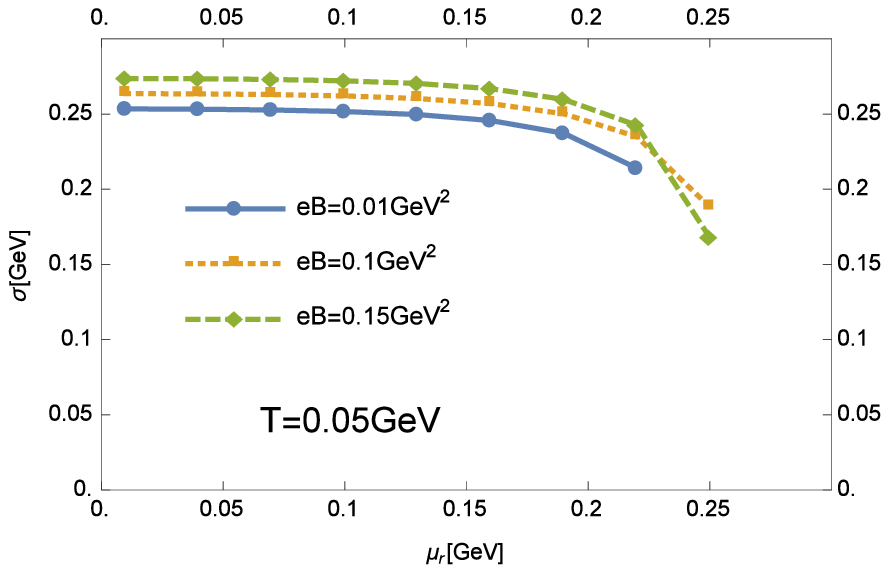}
 \includegraphics[width=3in]{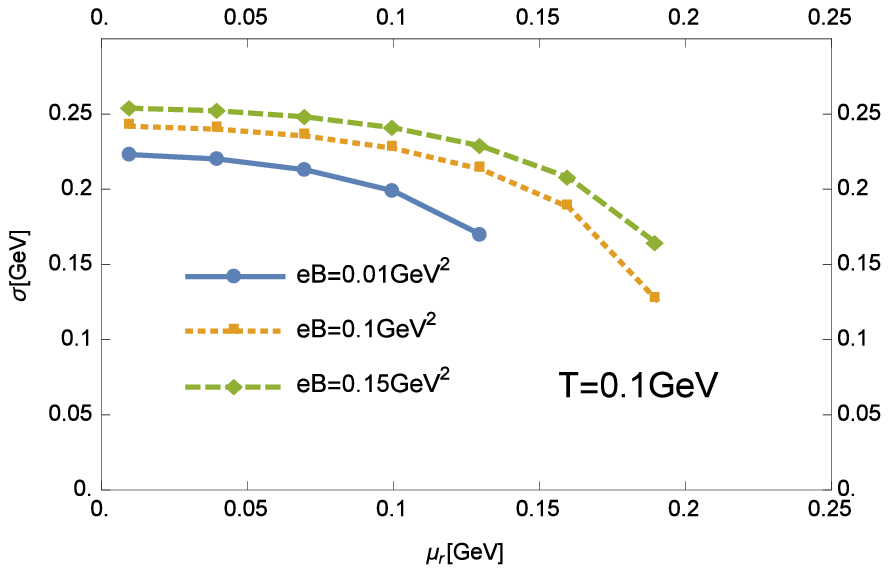}
 \includegraphics[width=3in]{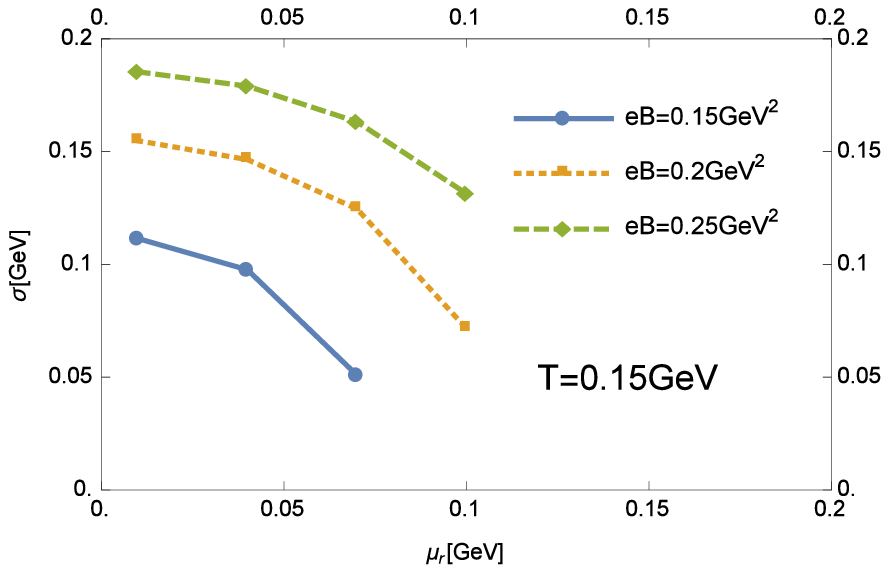}
 \caption{The $\mu_\text{r}$ dependance of dynamical mass $\sigma$ with fixed temperatures and different $eB$s.\label{ti}}
\end{figure}
\begin{figure}
 \centering
 \includegraphics[width=3in]{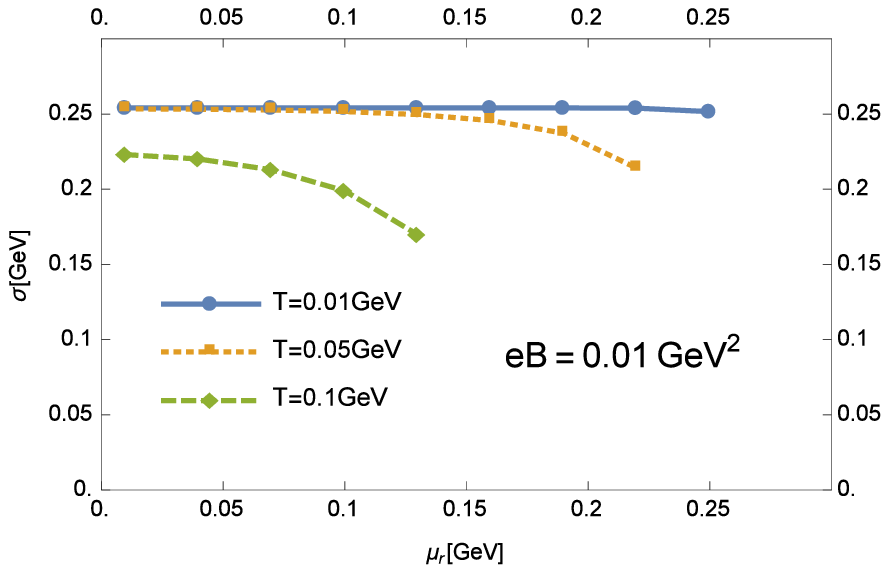}
 \includegraphics[width=3in]{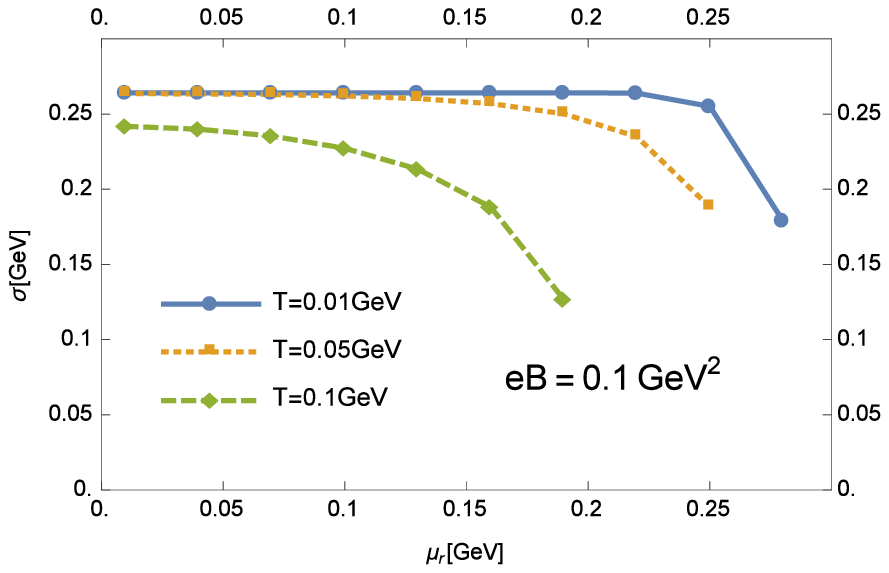}
 \includegraphics[width=3in]{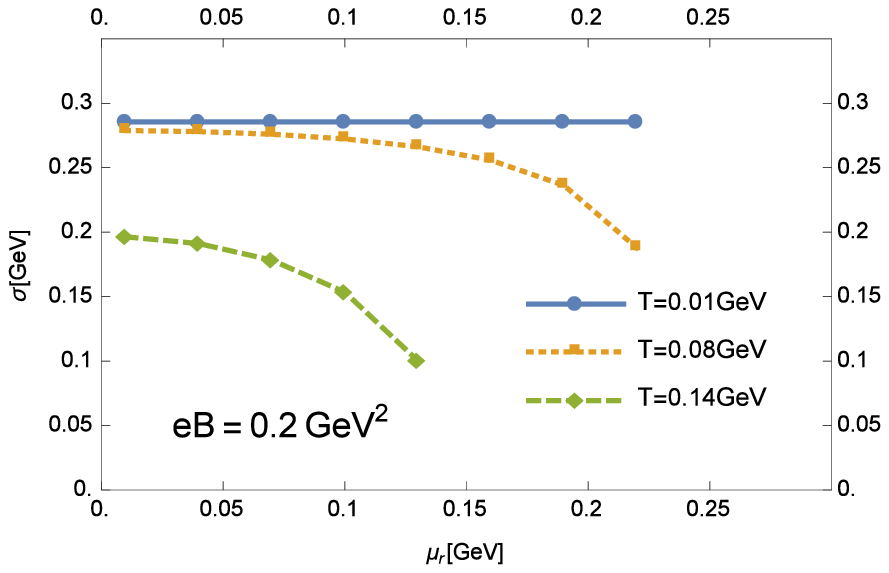}
 \includegraphics[width=3in]{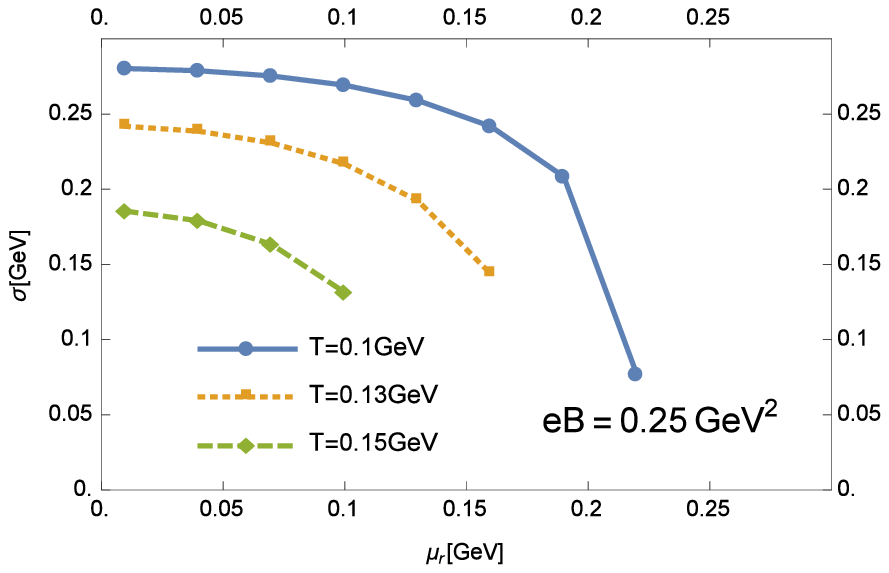}
 \caption{The $\mu_\text{r}$ dependance of dynamical mass $\sigma$ with fixed $eB$s and different temperatures.\label{ebi}}
\end{figure}

Now we employ Eqs. (\ref{sigma}) and (\ref{b}) with the cutoff of proper time $s$ to numerically calculate $\sigma$ and $b$, and pick out several representative results shown in Figs. (\ref{ti}) and (\ref{ebi}), clearly we can see when renormalized chemical potential reaches a critical point, phase transition happens, of course such critical point depends on temperature and magnetic field. In Fig. (\ref{ti}), the $\sigma$-$\mu_\text{r}$ relations just look like the classic results about condensate, when $\mu_\text{r}$ is smaller than a critical point, the state is in Nambu phase, and generally speaking, the stronger magnetic field and lower temperature is, the bigger dynamical mass in Nambu phase. But this is not a categorical conclusion with nonzero chemical potential, for example when $T=0.01$GGeV, comparing $eB=0.1\text{GeV}^{-1}$ line and $eB=0.15\text{GeV}^{-1}$ line, we can see that along with the increasing $\mu_\text{r}$, both lines start descending (this kind of descending normally can be seen as a second phase transition, but here we still count it as part of Nambu phase), in the middle of descending, inversely, the stronger magnetic field is, the smaller the generated dynamic mass we have, this is also some kind of `inverse magnetic catalysis effect', when $T=0.05$GeV there is the same effect. In Fig. (\ref{ebi}), the temperature dependance of dynamic mass is much simpler, higher temperature causes smaller dynamic mass, there is no `inverse' property. One more thing need to emphasize here, in Fig. (\ref{ti}), no matter what the specific temperature and magnetic field are, as soon as $\mu_\text{r}$ crosses a critical point, the state of Nambu phase will drop down to Wigner phase (zero dynamic mass).
\begin{figure}
 \centering
 \includegraphics[width=3in]{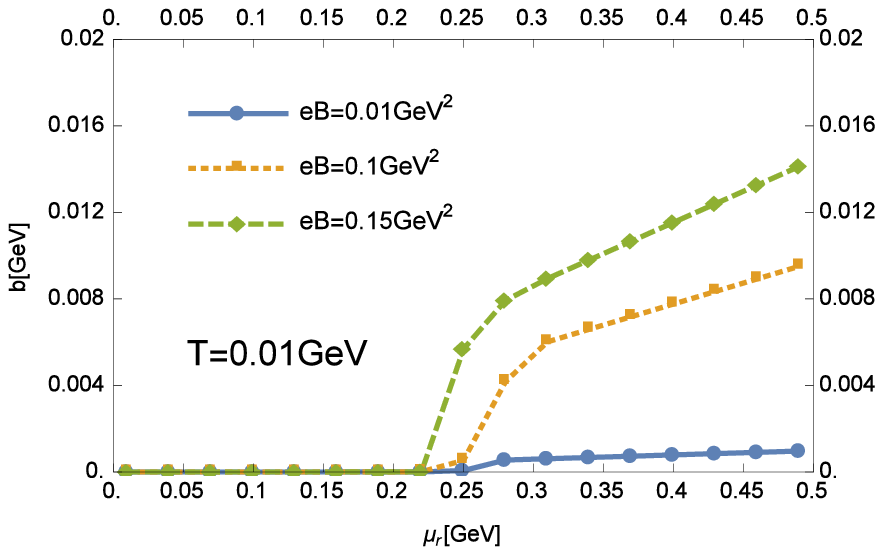}
 \includegraphics[width=3in]{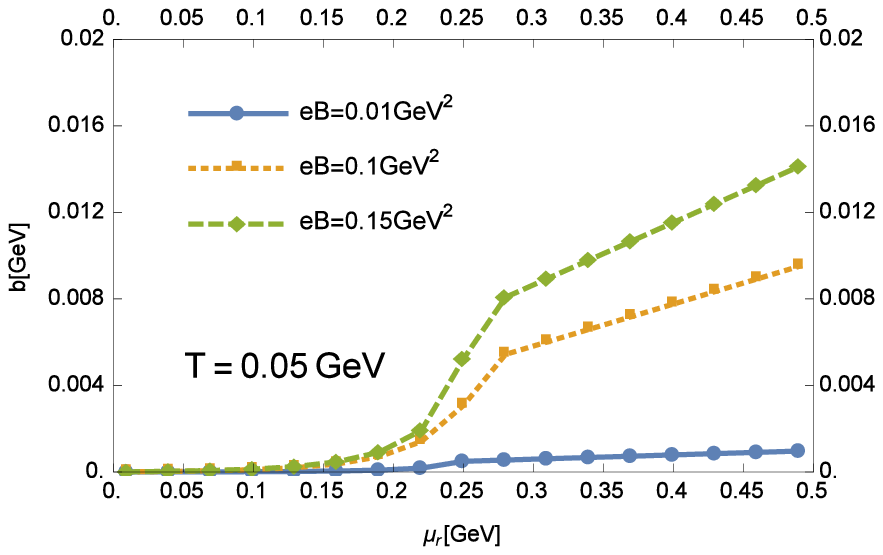}
 \includegraphics[width=3in]{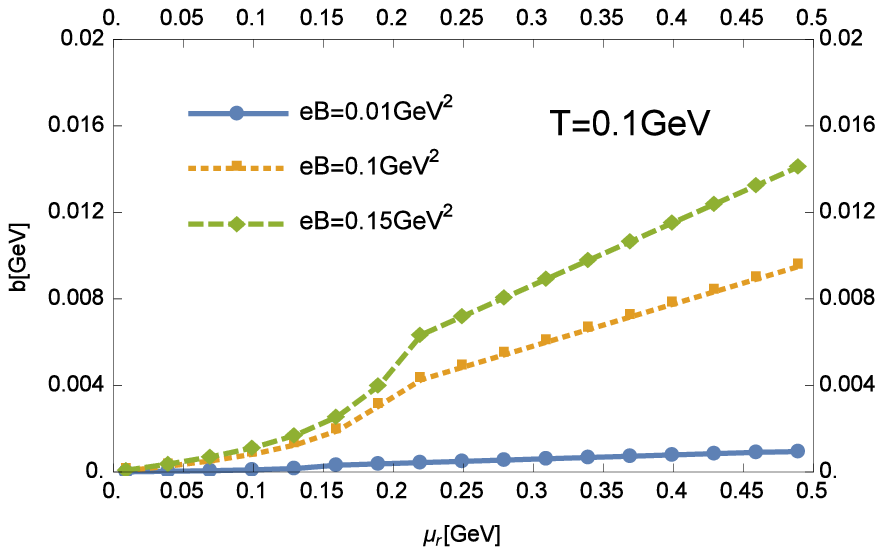}
 \includegraphics[width=3in]{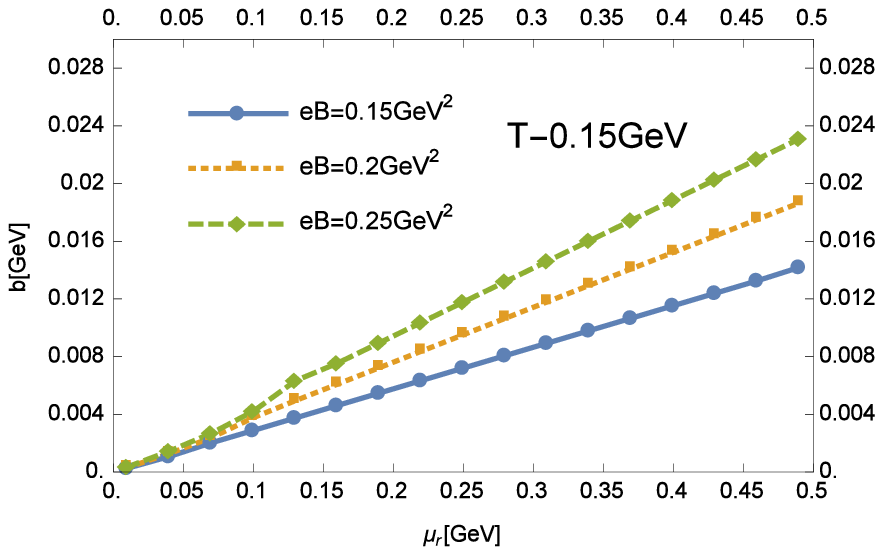}
 \caption{The $\mu$ dependance of $b$ with fixed temperatures and different $eB$s.\label{t1}}
\end{figure}
\begin{figure}
 \centering
 \includegraphics[width=3in]{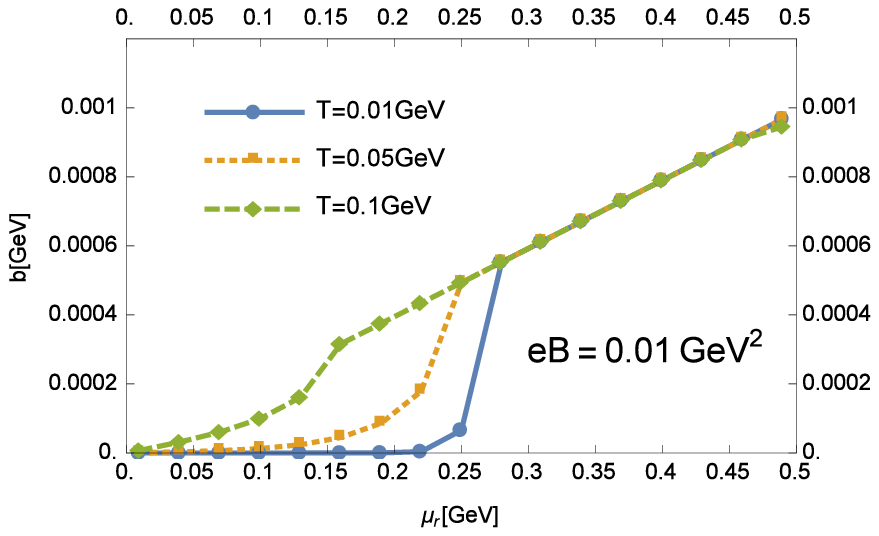}
 \includegraphics[width=3in]{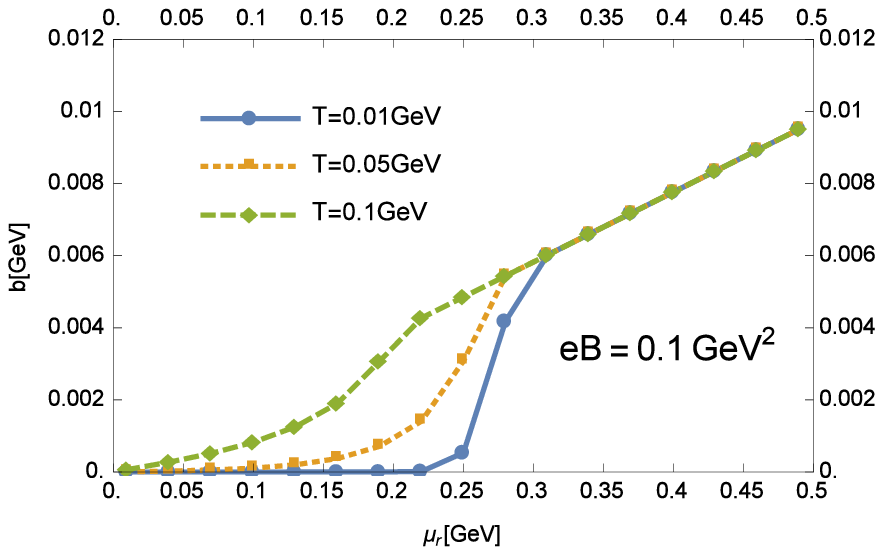}
 \includegraphics[width=3in]{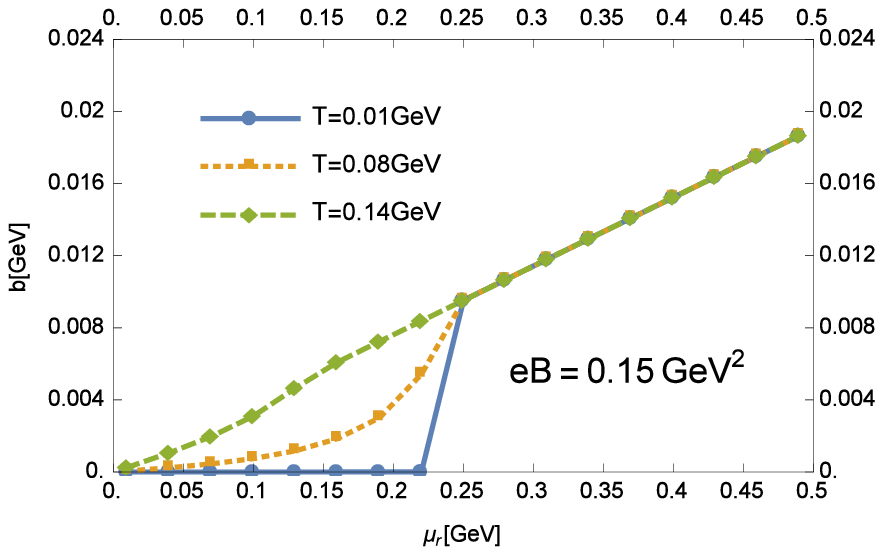}
 \includegraphics[width=3in]{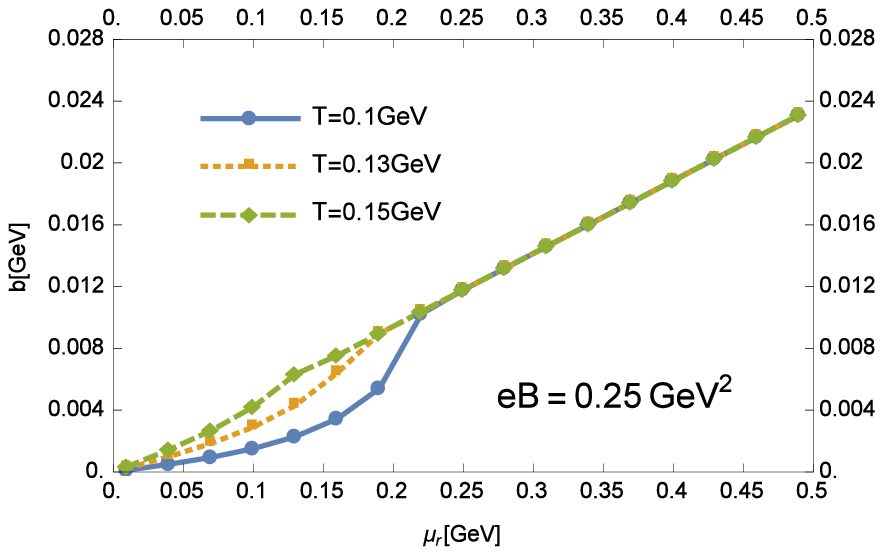}
 \caption{The $\mu$ dependance of $b$ with fixed $eB$s and different temperatures.\label{eb1}}
\end{figure}

Comparing with $\sigma$-$\mu_\text{r}$, the corresponding $b$-$\mu_\text{r}$ relations are shown in Figs. (\ref{t1}) and (\ref{eb1}). In these two figures, one can see that the gaps also appears at the same critical points (or critical areas) of $\mu_\text{r}$ as the corresponding ones in $\sigma$-$\mu_\text{r}$ do. When $\mu_\text{r}$ crosses these critical points, all $b$-$\mu_\text{r}$ relations are nearly straight lines. Similar to dynamical mass, with fixed temperature, the stronger magnetic field is, the bigger $b$ is, but for pseudovector condensate, there no such thing as `inverse magnetic catalysis effect', and with magnetic field fixed (Fig. (\ref{eb1})), higher temperature generally makes bigger $b$ when the state is in Nambu phase. Clearly the gaps of pseudovector condensate is smaller when temperature is higher.

Actually in Wigner phase, why $b$-$\mu_\text{r}$ relations act like straight lines can be answered by gap equations. In Wigner phase, dynamic mass is so small, we can treat it as zero in Eq. (\ref{b}), this gives us a simplified version of gap equation for pseudovector condensate,
\begin{equation}
 \frac{4\pi^2}{G}b=2eB\mu_\text{r},\label{b0}
\end{equation}
in which obviously $b$ and $\mu_\text{r}$ are linearly dependent. Noticing, this equation does not obviously depend on temperature.

In Nambu phase, the pseudovector condensate is too small to affect quark condensate, but why pseudovector condensate is so small? Actually we can find the solution in Eq. (\ref{b}), if we only consider the primary term of $b$ in the RHS of Eq. (\ref{b}), the gap equation for pseudovector condensate becomes
\begin{equation}
  \frac{4\pi^2}{G}b=eB\int\bigg[\frac{1}{1+e^{\beta(\omega-\mu_\text{r})}}-\frac{1}{1+e^{\beta(\omega+\mu_\text{r})}}\bigg]\,dp_3,\label{b3}
\end{equation}
this equation implies the main contribution to $b$ is the pure particle number in LLL (Lowest Landau Level), beside that, $b$'s existence depends on nonzero chemical potential and external magnetic field, these evidences support the conclusion that in a non-neutral system made of high energy particles, external magnetic field could stimulate weak pseudovector current (because $b$ is proportional to $\langle\bar\psi\gamma^5\gamma^3\psi\rangle$ from the definition in Eq. (\ref{mean})), and this pseudovector current is nearly proportional to pure particle number in LLL. $b$ is small due to the tiny difference between particle number and anti-particle number in LLL. What about the contribution from higher Landau levels, referring to Eq. (\ref{b}), it is $\int\sum_{n=1}^{+\infty}(F_{-n\text{f}}-F_{+n\text{f}})\,dp_3$, this is not pure particle number in higher Landau levels, and its value is much smaller than the contribution from Eq. (\ref{b3}).

\section{Conclusions and Remarks}

In this paper, we have studied the self-energy of NJL model with temperature, chemical potential and external magnetic field, turning out when chemical potential is nonzero, the self-energy is not only dynamic mass (or scalar condensate), it contains other condensates. In order to obtain correct gap equations in such case, one should have the original NJL model Fierz transformed firstly, then pick out full but raw self-energy from Lagrangian, and use symmetry analysis to simplify the self-energy. The treated self-energy (\ref{form}) contains four kinds of condensates, in this article we only study two of them (quark condensate $\sigma$ and pseudovector condensate $b$), the primary reason we exclude other two condensates is that the complete gap equations are too complex to be properly treated, but vector condensate $a$ can be absorbed into chemical potential, and tensor condensate $c$ is expected to have little effect on quark condensate, thus at qualitative level, two of the four condensates are adequate.

In Nambu phase, magnetic field strengthen quark condensate when dressed chemical potential $\mu_\text{r}$ is small enough, on this occasion, particles (or quarks) in the system are too sparse to interfere each other, therefore the $\sigma$ is insensitive to chemical potential, this is best supported by the nearly horizontal lines of three different magnetic fields in Fig. (\ref{ti}). When $\mu_\text{r}$ exceeds a threshold, particle density becomes important, interference among particles drastically increases, which leads to the descending of dynamic mass (the strength of scalar condensate) in $\sigma$-$\mu_\text{r}$ relation. Sometimes the descending is so sharp that one can treat this part as another phase, e.x. in Fig. (\ref{ti}), when $T=0.01$GeV, $eB=0.1$GeV$^2$ with $\mu_\text{r}$ ranging $0.25$-$0.30$GeV and $eB=0.15$GeV$^2$ with $\mu_\text{r}$ ranging $0.23$-$0.26$GeV, name as `intermediate phase', and it is likely a second phase transition to Nambu phase. During descending (the intermediate phase), the stronger magnetic field is, the faster dynamic mass drops, hence the smaller dynamic mass is generated, this is where the `inverse magnetic catalysis' happens, $T=0.01$GeV and $T=0.05$GeV in Fig. (\ref{ti}) are good illustration to such orientation. On the other hand, higher temperature will gradually smear the intermediate phase, the connection between it and Nambu phase becomes smooth, e.x. the intermediate phases of $T=0.1$GeV and $T=0.15$GeV are barely noticeable, one can also refer to Fig. (\ref{ebi}) for explicit impression of the smearing.

Now let us go on to the next phase, as $\mu_\text{r}$ keeps raising, it will eventually cross a critical point which indicates a threshold of first phase transition, then dynamic mass jumps to Wigner phase ($\sigma=0$).

From the diagrams of $b$-$\mu_\text{r}$ relation, we can see pseudovector condensate $b$ also have gaps between Nambu phase and Wigner phase, but most of the time the gaps are not as obvious as dynamic mass is, and beside that, $b$ are very small in Nambu phase, therefore pseudovector condensate is not a good order parameter for phase transition. pseudovector condensate depends on two conditions, nonzero external magnetic field and surplus charges in the whole system. For a charge neutral system, its total magnetic moment is zero, so its quantum states are degenerate, while if the system has surplus charges, an external magnetic field will certainly stimulate nonzero magnetic moment of this system, the degeneracy of quantum states is removed. The pseudovector condensate is similar to such magnetic moment, and we would like to find out its relation with magnetic moment in following works. Because of the magnetic-moment-like property, it is quite reasonable $b$ can be strengthened by magnetic field and chemical potential, and inversely, the strengthening is also a convincing evidence that pseudovector condensate is magnetic-moment-like. We are still working on explaining the mechanism that hides behind the puzzling relation between temperature and pseudovector condensate both in Nambu phase and quasi-Wigner phase. In later investigation we also find out that with the limitation of $\sigma=0$, Eq. (\ref{b0}) could relate to `chiral separation effect (CSE)',
\begin{equation}
 \vec j_\text{A}=N_\text{c}\sum_\text{f}\frac{q_\text{f}^2\mu}{2\pi^2}\vec B, \label{cse}
\end{equation}
which was first discovered by chiral anomaly \cite{metli,fukushima2,vil,gio,fukushima3}, comparing these two equations, they are quite similar except little factor difference, and in CSE, the chiral anomaly is irrelevant to temperature \cite{shi2}, this mechanism could explain why pseudovector condensate prefers not responding to temperature. Nevertheless, when dynamical mass approaches zero, $b$ is nearly only proportional to magnetic field and dressed chemical potential, which brings us two meaningful results: Firstly, self-energy is no longer trivial even in Wigner phase, although dynamical mass is zero, pseudovector condensate is nothing close to zero. Secondly, if chemical potential or magnetic field is large enough, pseudovector condensate could be strong enough to produce obvious effects. And we think more importantly, according to Eq. (\ref{dispersion}), pseudovector condensate causes splitting of dispersion relation. Especially in Wigner phase, $b$ is capable of acquiring bigger values, the dispersion relation Eq. (\ref{dispersion}) can be rewritten as
\begin{equation}
 \omega_{\pm n\text{f}}=\sqrt{(|p_3|\pm b)^2+2n|q_\text{f}|eB},\label{dispersion2}
\end{equation}
it is shown that at specific momentum and Landau level, energy level splits, a hidden degeneracy is removed.

Physical effects induced by pseudovector condensate should depend on dispersion relation, the degeneracy removing in Eq. (\ref{dispersion2}) causes different particle number densities. In a series of Tatsumi's works \cite{tatsumi1, tatsumi2, tatsumi3}, they use $b$ to study ferromagnetism in nuclear matter and quark matter. $b$, as a condensate of pseudovector current, is the parameter to describe `spin polarization' in their articles, it is said that $b$'s existence slightly splits dispersion relation, hence the `spin'-up particles and `spin'-down particles are separated to different Fermi surfaces, known as `spin polarization'. Different to our study, they does not consider external magnetic field in their works, therefore $b$ will not exist until quarks are massive and the quark matter is in CSC (color super conductivity) state. As soon as quark matter leaves CSC state, `spin polarization' vanishes as $b$ does, the `spin polarization' effect is spontaneous. However, in our article, the fermion's original mass is zero (chiral limit), and CSC state is not included, but we have nonzero external magnetic field, the magnetic field keeps $b$ presenting, hence `spin polarization' is automatically but not spontaneously stimulated. In this article, the strength of external magnetic field ranges $0.01$-$0.25$GeV$^2$ ($10^{19}$-$10^{20}$ Gauss), this is the strongest magnetic field could be found in experiments, even so, the splitting is too small to be distinguished from background in the experiments. We are looking forward to other conditions that can produce obvious effects, e.x. large space scale could enlarge tiny modification in dispersion relation, the magnetar, a kind of neutron star, is expected to have magnetic field of $10^{15}$ Gauss, although this strength is several order of magnitudes smaller than what we have considered in this paper, magnetars have large volumes, it could enlarge the splitting in energy levels. On the other hand, we can also simply expect larger magnetic field to produce obvious effects in future experiments. It is believed that in early universe, the magnetic field could reach $10^{23}$ Gauss, the results we have may play a role in explaining universe evolution.

The inverse magnetic catalysis effect mentioned in this article is not different to the well-known `inverse magnetic catalysis' studied in these articles \cite{bali,bruckmann}, with lattice QCD, they show inverse magnetic catalysis when $\mu=0$ and $T\neq0$. In NJL model, this kind of effect can not be achieved by mean field approximation at least. We are looking forward to use other methods beyond mean field approximation to find out the well-known inverse magnetic catalysis in NJL model. Pseudovector condensate is irrelevant to the inverse magnetic catalysis we discuss in this article, the primary factor is chemical potential, or particle density.

\appendix
\section{The Deduction of Gap Equations}
\label{sec:append}
Through Fierz transformation and mean field approximation, we are able to acquire Lagrangian $\mathcal{L}'$ in Eq. (\ref{new}). In order to deduce the gap equations at finite temperature and chemical potential, we rewrite $\mathcal{L}'$ as
\begin{equation}
 \mathcal{L}'=\bar\psi(-\gamma^0\frac{\partial}{\partial\tau}+\gamma^i\hat\Pi_i-\Sigma)\psi+\mathcal{L}_\text{M}+\mu\bar\psi\gamma^0\psi,\label{new2}
\end{equation}
the partition function (only has the functional integral of fermion field) is
\begin{equation}
 \mathcal{Z}=\int D\bar\psi D\psi\,e^{\int_0^\beta d\tau\int d\vec{x}\,\mathcal{L}'}=e^{\mathcal{W}[\sigma, a, b, c]},\qquad\beta=\frac{1}{T},
\end{equation}
\begin{equation}
 \mathcal{W}[\sigma, a, b, c]=\mathcal{L}_\text{M}\int_0^\beta d\tau\int d\vec{x}+N_\text{c}\Tr_\text{sf}\ln\big(-\gamma^0\frac{\partial}{\partial\tau}+\gamma^i\hat\Pi_i-\Sigma+\mu\gamma^0\big), \label{eff}
\end{equation}
the trace operator `$\Tr_\text{sf}$' in the effective action Eq. (\ref{eff}) implies summing up expectation values of $\ln(/\kern-0.55em\hat\Pi-\Sigma+\mu\gamma^0)$ at all quantum state and tracing the matrices in both flavor and spinor spaces. The flavor space is not trivial because of $Q$ from $\hat\Pi_{1,2}$ (u quark and d quark have different electric charges), but $Q$ is diagonal, we can separate flavor space apart. Beside this, in the effective action, $\mu$ and $a$ from $\Sigma$ can combine to a new chemical potential $\mu_\text{r}=\mu-a$ (named as `renormalized chemical potential'), therefore we can rewrite Eq. (\ref{eff}) as
\begin{equation}
 \mathcal{W}[\sigma, a, b, c]=\mathcal{L}_\text{M}\beta\int d\vec{x}+N_\text{c}\sum_\text{f}\Tr_\text{s}\ln(\hat S_\text{f}^{-1}), \label{eff2}
\end{equation}
\begin{equation}
 \hat S_\text{f}=(/\kern-0.55em\hat\Pi^\text{f}-\tilde\Sigma-c\sigma^{12})^{-1},
 \qquad\hat\Pi^\text{f}_0=-\frac{\partial}{\partial\tau}+\mu_\text{r}=\hat p_0\qquad\hat\Pi^\text{f}_i
 =\hat p_i+q_\text{f}eA_\mu\qquad\tilde\Sigma=\sigma+b\gamma^5\gamma^3,
\end{equation}

The gap equations are partial differentiations of effective action with its variables $\sigma$, $a$, $b$ and $c$,
\begin{equation}
 \frac{\delta\mathcal{W}}{\delta\sigma}=0,\qquad\frac{\delta\mathcal{W}}{\delta a}=0,\qquad\frac{\delta\mathcal{W}}{\delta b}=0,\qquad\frac{\delta\mathcal{W}}{\delta c}=0,\label{gap0}
\end{equation}
in more explicit forms, we have
\begin{equation}
 \frac{4}{4N_\text{c}+1}\frac{\sigma}{G}\int d\vec{x}=-T\sum_\text{f}\Tr_\text{s}\hat S_\text{f}, \label{gap11}
\end{equation}
\begin{equation}
 \frac{2}{G}a\int d\vec{x}=T\sum_\text{f}\Tr_\text{s}(\hat S_\text{f}\gamma^0), \label{gap12}
\end{equation}
\begin{equation}
 \frac{2}{G}b\int d\vec{x}=-T\sum_\text{f}\Tr_\text{s}(\hat S_\text{f}\gamma^5\gamma^3), \label{gap13}
\end{equation}
\begin{equation}
 \frac{4}{G}c\int d\vec{x}=-T\sum_\text{f}\Tr_\text{s}(\hat S_\text{f}\sigma^{12}), \label{gap14}
\end{equation}
in these equations, we need to deal with `$\Tr_\text{s}$' (summing all expectation values of $\hat S_\text{f}$ and tracing gamma matrices), in order to do that, firstly we need to make $\hat S_\text{f}$ more convenient to calculate, so here it is
\begin{eqnarray}
 \hat S_\text{f}&=&\frac{1}{/\kern-0.55em\hat\Pi^\text{f}-\tilde\Sigma-c\sigma^{12}}
 =\frac{/\kern-0.55em\hat\Pi^\text{f}+\tilde\Sigma-c\sigma^{12}}
 {(/\kern-0.55em\hat\Pi^\text{f}-c\sigma^{12})^2-\tilde\Sigma^2+[/\kern-0.55em\hat\Pi^\text{f}-c\sigma^{12},\tilde\Sigma]}\nonumber\\
 &=&\frac{/\kern-0.55em\hat\Pi^\text{f}+\tilde\Sigma-c\sigma^{12}}{\hat p_0^2-(\hat\Pi^\text{f}_\perp)^2-\hat p_3^2+2b\hat p_3\gamma^5+2c\hat p_0\gamma^5\gamma^3-M},\label{prop}
\end{eqnarray}
\begin{equation}
 (\hat\Pi^\text{f}_\perp)^2=(\hat\Pi^\text{f}_1)^2+(\hat\Pi^\text{f}_2)^2,\qquad M=\sigma^2+b^2-c^2+q_\text{f}eB\sigma^{12}+2\sigma b\gamma^5\gamma^3.
\end{equation}
as we can see in Eq. (\ref{prop}), rewriting $\hat S_\text{f}$ does not scalarize the denominator, but it implies a new set of operators $(\frac{\partial}{\partial\tau}, (\hat\Pi^\text{f}_\perp)^2, \hat p_3)$ that commute with each other. We know $\Tr_\text{s}$ is representation irrelevant, thus we may introduce the eigenstates of $(\frac{\partial}{\partial\tau}, (\hat\Pi^\text{f}_\perp)^2, \hat p_3)$, and quantize the denominator of $\hat S_\text{f}$. Such eigenstate is defined as $|m;n,\lambda;p_3\rangle=|m\rangle_0\otimes|n,\lambda\rangle_{1,2}\otimes|p_3\rangle_3$ (the indexes 0,1,2,3 represent the Hilbert spaces that relate to $\{x^0$,$x^1$,$x^2$,$x^3\}$, we would ignore these indexes below if there did not cause any misunderstanding), it satisfies
\begin{equation}
 \frac{\partial}{\partial\tau}|m\rangle=i\omega_m|m\rangle, \qquad\omega_m=(2m+1)\pi T,  \qquad m\in\mathbb{Z},
\end{equation}
\begin{equation}
 (\hat\Pi^\text{f}_\perp)^2|n,\lambda\rangle=(2n+1)|q_\text{f}|eB|n,\lambda\rangle, \qquad n\in\mathbb{N}^0,
\end{equation}
\begin{equation}
 \hat p_3|p_3\rangle=p_3|p_3\rangle, \qquad p_3\in\mathbb{R},
\end{equation}
in $|n,\lambda\rangle$, $\lambda$ is a free variable that ranges all real numbers, and $|n,\lambda\rangle$ is normalized so we have $\langle n',\lambda'|n,\lambda\rangle=\delta_{nn'}\delta(\lambda-\lambda')$. One can refer to our previous work \cite{shi} for more detailed description about $|n,\lambda\rangle$ state. By introducing eigenstate, the denominator of $\hat S_\text{f}$ can be completely quantized, taking $\Tr_\text{s}\hat S_\text{f}$ as example, there is
\begin{equation}
 \Tr_\text{s}\hat S_\text{f}=\sum_{m=-\infty}^{+\infty}\sum_{n=0}^{+\infty}\int_{-\infty}^{+\infty}d\lambda\int_{-\infty}^{+\infty}dp_3\,
 \frac{\langle m;n,\lambda;p_3|(/\kern-0.55em\hat\Pi^\text{f}+\tilde\Sigma-c\sigma^{12})|m;n,\lambda;p_3\rangle}
 {p_0^2-(2n+1)|q_\text{f}|eB-p_3^2+2bp_3\gamma^5+2cp_0\gamma^5\gamma^3-M},\label{prop2}
\end{equation}
\begin{equation}
 p_0=-i\omega_m+\mu_\text{r}.
\end{equation}
in Eq. (\ref{prop2}), the numerator of integrand is not completely quantized, because $\hat\Pi^\text{f}_{1,2}$ are not eigen-operators of $|m;n,\lambda;p_3\rangle$, but we will prove the contribution from $\hat\Pi_{1,2}$ in the numerator is zero after the integral of $\lambda$ and $p_3$. Here taking $\hat\Pi_1$ as an example, referring to the method in \cite{shi}, we introduce a complete state $|\Pi_1,p_1\rangle_{1,2}$ which is the eigenstate of $(\hat\Pi_1^\text{f},\hat p_1)$, and there is
\begin{equation}
 \langle\Pi_1,p_1|n,\lambda\rangle=c_ne^{i\lambda(p_1+\frac{\Pi_1}{2})}h_n(\sqrt{\frac{2}{|q_\text{f}|eB}}\Pi_1),
\end{equation}
$h_n$ is the solution of Weber differential equation, it is an even function of $\Pi_1$, thus when we insert projection operator $\int d\Pi_1dp_1\,|\Pi_1,p_1\rangle\langle\Pi_1,p_1|$ into the numerator, we can prove
\begin{equation}
 \int d\lambda\,\langle n,\lambda|\int d\Pi_1dp_1\,|\Pi_1,p_1\rangle\langle\Pi_1,p_1|\hat\Pi^\text{f}_1|n,\lambda\rangle
 \propto\int d\lambda d\Pi_1dp_1\,\Pi_1h_n^2(\sqrt{\frac{2}{|q_\text{f}|eB}}\Pi_1)=0£¬
\end{equation}
analogously $\hat\Pi_2$'s contribution is also zero.

Next, we want to scalarize the denominator of $\hat S_\text{f}$, and write it in a form of linear combination of sixteen Dirac matrices $\{I_4,\gamma^\mu,\gamma^5,\gamma^5\gamma^\mu,\sigma^{\mu\nu}\}$, this is barely accessible manually, but through Mathematica programs, we are able to find out the combination (especially the coefficients that couple with Dirac matrices). It turns out some of Dirac matrices are missing in the combination (the corresponding coupling coefficients are zero), finally we are able to get an effective denominator-scalarized form with the combination of $\{I_4,\gamma^0,\gamma^5\gamma^3,\sigma^{12}\}$ in $\hat S_\text{f}$, wrote as
\begin{equation}
 S_\text{eff}=f_1I_4+f_2\gamma^0+f_3\gamma^5\gamma^3+f_4\sigma^{12}. \label{seff}
\end{equation}
here the summation of all $(2n+1)|q_\text{f}|eB$ (Landau levels) terms have already contained in $f_1$, $f_2$, $f_3$ and $f_4$, leaving sum of $\omega_m$ and integral of $p_3$ outside $S_\text{eff}$.

Consequently the gap equations are
\begin{equation}
 \frac{2}{4N_\text{c}+1}\frac{\sigma}{G}=-T\frac{|q_\text{f}|eB}{\pi^2}\sum_\text{f}\sum_m\int f_1\,dp_3, \label{sigmaf1}
\end{equation}
\begin{equation}
 \frac{a}{G}=T\frac{|q_\text{f}|eB}{\pi^2}\sum_\text{f}\sum_m\int f_2\,dp_3, \label{af2}
\end{equation}
\begin{equation}
 \frac{b}{G}=-T\frac{|q_\text{f}|eB}{\pi^2}\sum_\text{f}\sum_m\int f_3\,dp_3. \label{bf3}
\end{equation}
\begin{equation}
 \frac{2}{G}c=-T\frac{|q_\text{f}|eB}{\pi^2}\sum_\text{f}\sum_m\int f_4\,dp_3. \label{cf4}
\end{equation}

The complete expression of $\{f_1,f_2,f_3,f_4\}$ are quite complicate, therefore in this article we try to make some simplification to the gap equations above. Firstly, we take off Eq. (\ref{af2}) from gap equations, as we can see, parameter `$a$' is absorbed by $\mu_\text{r}$ in these equations (except in Eq. (\ref{af2})), and in our following studies, we treat $\mu_\text{r}$ as a free variable, hence there is no need for computing $a$'s value explicitly. Secondly, the presence of $b$ and $c$ makes equations hard to simplify at finite temperature, for example, the expressions of $\{f_1,f_2,f_3,f_4\}$ are rational functions with denominators like
\begin{equation}
 p_0^2+c^2-2n|q_\text{f}|eB-p_3^2-\sigma^2-b^2\pm2\sqrt{(cp_0-\sigma b)^2+b^2p_3^2},\label{de}
\end{equation}
obviously $p_0=-i\omega_m+\mu_\text{r}$ presents in a radical expression, this causes gap equations too complicate to calculate, hence we need some approximation to simplify equations. On the other hand, the values of $a$, $b$ and $c$ are expected very small (and in fact they are), in order to take $p_0$ out of the radical expression in Eq. (\ref{de}), a straight simplification is to set $b$ or $c$ zero. In this article, our main purpose is to evaluate phase transitions under the interference of pseudovector condensate $\langle\bar\psi\gamma^5\gamma^3\rangle$, therefore we assume $c=0$ (no tensor condensate) and exclude Eq. (\ref{cf4}), this left us only Eqs. (\ref{sigmaf1}) and (\ref{bf3}) to study. In conclusion, we only need the expression of $f_1$ and $f_3$
\begin{equation}
 f_1=\frac{\sigma}{4\omega}\bigg(\frac{1}{p_0-\omega_-}-\frac{1}{p_0+\omega_+}\bigg)
 +\frac{\sigma}{2\omega}\sum_\pm\sum_{n=1}^{+\infty}\frac{\omega_\pm}{p_0^2-\omega_{\pm n\text{f}}^2},\label{f1}
\end{equation}
\begin{equation}
 f_3=-\frac{1}{4}\bigg(\frac{1}{p_0-\omega_-}+\frac{1}{p_0+\omega_+}\bigg)
 +\frac{1}{2}\sum_{n=1}^{+\infty}\bigg(\frac{\omega_+}{p_0^2-\omega_{+n\text{f}}^2}-\frac{\omega_-}{p_0^2-\omega_{-n\text{f}}^2}\bigg).\label{f3}
\end{equation}
\begin{equation}
 \omega=\sqrt{p^2_3+\sigma^2},\quad\omega_\pm=\omega\pm b,\quad\omega_{\pm n\text{f}}=\sqrt{\omega^2_\pm+2n|q_\text{f}|eB},\quad n\in\mathbb{N}^0,\label{dispersion}
\end{equation}

\acknowledgments

This work is supported in part by the National Natural Science Foundation of China (under Grants No. 11275097, No. 11475085, and No. 11535005), the Advanced Research Foundation of National University of Defence Technology (under Grant No. ZK17-03-16).

\end{document}